\documentclass[sigconf]{acmart}

\usepackage{balance}
\usepackage{url}
\usepackage{tcolorbox}
\usepackage{enumitem}
\usepackage{xcolor}
\usepackage{multirow}
\definecolor{awesome}{RGB}{70,130,180}
\usepackage{hyphenat}

\newcommand{\figref}[1]{Figure~\ref{#1}}
\newcommand{\code}[1]{\textit{``#1''}}

\AtBeginDocument{%
  }

\copyrightyear{2026}
\acmYear{2026}
\setcopyright{cc}
\setcctype{by}
\acmConference[MSR '26]{23rd International Conference on Mining Software Repositories}{April 13--14, 2026}{Rio de Janeiro, Brazil}
\acmBooktitle{23rd International Conference on Mining Software Repositories (MSR '26), April 13--14, 2026, Rio de Janeiro, Brazil}
\acmPrice{}
\acmDOI{10.1145/3793302.3793612}
\acmISBN{979-8-4007-2474-9/2026/04}

\begin{document}

\title
[Who Writes the Docs in SE~3.0? Agent vs.\ Human Documentation Pull Requests]
{%
  Who Writes the Docs in SE~3.0?\\
  \vspace{0.1em}
   Agent vs.\ Human Documentation Pull Requests%
}

\author{
Kazuma Yamasaki\textsuperscript{$\dagger$},
Joseph Ayobami Joshua\textsuperscript{$\dagger$},
Tasha Settewong\textsuperscript{$\dagger$}\\
Mahmoud Alfadel\textsuperscript{$\ddagger$},
Kazumasa Shimari\textsuperscript{$\dagger$},
Kenichi Matsumoto\textsuperscript{$\dagger$}
}

\affiliation{%
  \institution{\textsuperscript{$\dagger$}Nara Institute of Science and Technology, Japan}
  \country{}
}

\affiliation{%
  \institution{\textsuperscript{$\ddagger$}University of Calgary, Canada}
  \country{}
}

\email{{yamasaki.kazuma.yj9, joseph.ayobami_joshua.je3,tasha.settewong.ts1}@naist.ac.jp}
\email{mahmoud.alfadel@ucalgary.ca, {k.shimari, matumoto}@is.naist.jp}

\begin{abstract}
As software engineering moves toward SE~3.0, AI agents are increasingly used to carry out development tasks and contribute changes to software projects. It is therefore important to understand the extent of these contributions and how human developers review and intervene, since these factors shape the risks of delegating work to AI agents. 
While recent studies have examined how AI agents support software development tasks (e.g., code generation, issue resolution, and PR automation), their role in documentation tasks remains underexplored--even though documentation is widely consumed and shapes how developers understand and use software.

Using the AIDev, we analyze 1,997 documentation-related pull requests (PRs) authored by AI agents and human developers, where documentation PRs are those that create or modify project documentation artifacts.
We find that AI agents submit substantially more documentation-related PRs than humans in the studied repositories. 
We further observe that agent-authored documentation edits are typically integrated with little follow-up modification from humans, raising concerns about review practices and the reliability of agent-generated documentation. 
Overall, while AI agents already contribute substantially to documentation workflows, our results suggest concerns for emerging challenges for documentation quality assurance and human--AI collaboration in SE~3.0.

\end{abstract}



\maketitle

\renewcommand{\shortauthors}{Kazuma Yamasaki, Joseph Ayobami Joshua, Tasha Settewong, Mahmoud Alfadel, Kazumasa Shimari, Kenichi Matsumoto}

\section{Introduction}

SE~3.0~\cite{hassan2024towards} marks a shift toward software development environments in which autonomous AI agents act as teammates that can propose and implement changes, rather than passive assistants that only respond to prompts.
Recent work highlights this transition and the growing capabilities of agentic tools
to generate and submit project updates through typical collaboration mechanisms such as pull requests (PRs)~\cite{hassan2024towards,watanabe2025use}.
Industry reports similarly suggest rapid adoption of AI assistance in practice~\cite{stackoverflow_survey_2024_ai}.

To quantify agents' participation in development workflows and assess their impact, prior work has largely focused on coding-centric activities~\cite{chen2021evaluating,xia2023aprllm,watanabe2025use}.
However, we still know little about agent involvement in documentation---a critical artifact for understanding, onboarding, maintenance, and reuse---and how human developers respond to such contributions in real PR workflows.
MSR research has extensively studied documentation artifacts and quality issues~\cite{ikeda2019empirical,tang2023evaluating,codabux2021technical}, but evidence on agent-authored documentation in practice remains limited.

Prior work on documentation generation by large language models (LLMs) has mostly evaluated output quality in controlled settings (e.g., correctness, usefulness, and readability)~\cite{macke2024testing}.
In contrast, agent-authored documentation changes occur within real collaboration workflows, where integration, review, and subsequent revision determine whether documentation is trusted and maintained.
This setting raises two practical questions: (i) how prevalent agent-authored documentation changes are relative to human-authored ones, and (ii) whether documentation changes proposed by agents receive meaningful human follow-up after being integrated.

To answer these questions, we analyze documentation-related PRs in the AIDev~\cite{watanabe2025use} and compare agent-authored and human-authored contributions.
Specifically, we structure our study around the following research questions:

\begin{enumerate}[]

\item \textbf{RQ$_1$  (Prevalence): How many documentation-related PRs are created by AI agents compared to humans?}\\
Of the 1,997 documentation-related PRs, 1,478 are agent-authored and 519 are human-authored.
At the file level, 66.1\% of changed files are edited only by agents, 30.2\%  only by humans, and 3.7\%  are edited by both.

\item \textbf{RQ$_2$ (Integration): How often are agent-authored documentation PRs accepted?}\\
We find that an average of 86.8\% of line edits made by agents in source code are accepted by human developers. Among accepted agent edits, human follow-up is often limited: in 85.7\% of cases, agent additions exceed subsequent human deletions on the same files, and 34.5\% of cases show zero human deletions after agent additions.
\end{enumerate}

These results demonstrate that documentation has become a primary entry point for AI agent contributions, yet integrating agent-authored documentation introduces reliability risks that are not automatically mitigated by existing review practices. This study lays the foundation for future work on strengthening documentation quality assurance in SE~3.0 and on designing human--AI workflows and tools that better support accountable review and maintenance of agent-generated documentation.

We release a replication package to support reproducibility~\footnote{\url{https://github.com/NAIST-SE/msr2026-docs-prs-replication}}.
Our replication package includes (i) the derived analysis subset, (ii) scripts to reproduce the subset from the original dataset, and (iii) scripts to reproduce all analyses and figures.

\section{Existing Research}

\subsection{Software Documentation in Modern Repositories}

Software documentation is a critical artifact for understanding, onboarding, maintenance, and reuse, yet it is fragile and costly to maintain.
Aghajani et al.\ mined 878 documentation\hyp{}related artifacts from mailing lists, issue trackers, Stack Overflow, and PRs, producing a taxonomy of 162 documentation issue types.
They report common problems such as outdated, incomplete, and inconsistent information, and conclude that developers and users prefer documentation that is correct, complete, up-to-date, usable, maintainable, readable, and useful~\cite{10.1109/ICSE.2019.00122}.
Prana et al.\ analyzed 4{,}226 README sections from 393 GitHub repositories and proposed a content taxonomy, showing that many projects lack basic information about purpose and status despite rich usage instructions~\cite{prana2019readme}.
Work on code--comment co-evolution further shows that comments often lag behind code changes, causing inconsistency and confusion and motivating continuous documentation maintenance~\cite{wen2019commentevolution}.

Together, this literature underscores that documentation is critical yet fragile,
and that understanding documentation work in repositories remains an important MSR problem.
In this work, we study documentation\hyp{}related PRs, with an emphasis on documentation edits made by AI agents alongside humans.

\subsection{LLM-Based Documentation Editing in Pull Requests}

LLM-based tools are increasingly used for documentation work in practice.
Observational evidence from software engineering practice shows that developers use ChatGPT for tasks beyond code generation, including improving documentation wording and related documentation work~\cite{khojah2024beyond}.
Repository-scale evidence also suggests that documentation is a common task type for AI teammates: AIDev reports that documentation-related PRs account for a non-trivial share of agentic PRs and often exhibit comparatively high acceptance rates~\cite{li2025aidev}.
Moreover, MSR studies of ChatGPT usage in PR workflows indicate that developers use ChatGPT for purposes beyond code generation, including documentation refinement~\cite{ogenrwot2025patchtrack}.

Despite these signs of widespread use, we still lack a PR-level understanding of how LLM-driven documentation edits are carried out in real repositories.
Prior work does not isolate documentation\hyp{}focused PRs nor quantify how documentation changes are packaged and reviewed.
Our study addresses this gap by focusing specifically on documentation\hyp{}related PRs and by quantifying, at scale, the structure and outcomes of documentation edits authored by AI agents and humans.

\section{Approach}

\subsection{Data Collection}

\begin{figure*}[ht]
  \centering
  \includegraphics[width=\textwidth]{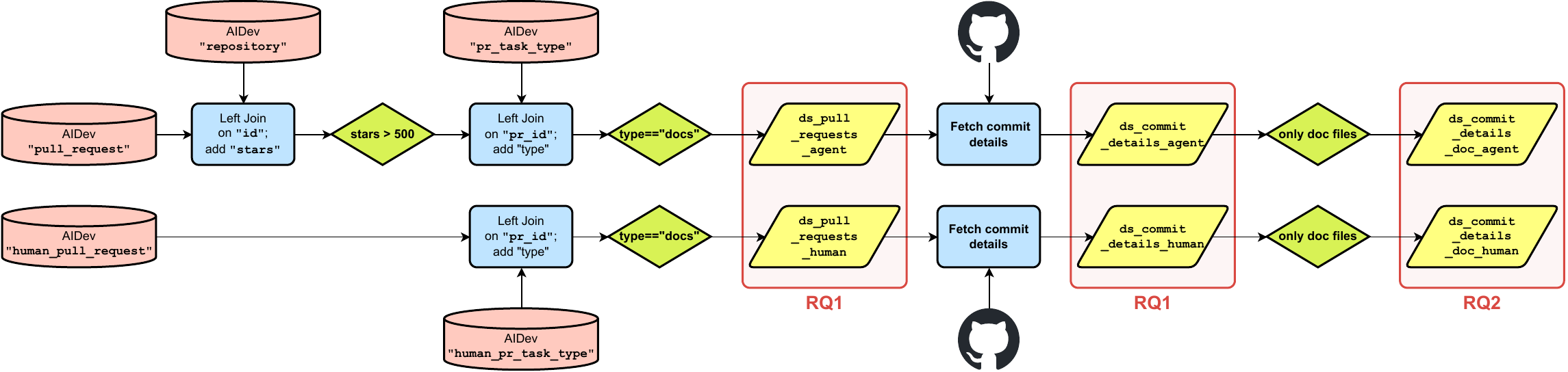}
  \caption{Flowchart of Data Collection}
  \label{fig:flowchart}
\end{figure*}

We construct an RQ-specific subset by extracting Agentic-PRs and Human-PRs that involve documentation changes from the AIDev~\cite{li2025aidev}, and then retrieving detailed commit-level information for each PR via GitHub API.
\figref{fig:flowchart} shows the data extraction process used in our analysis.

We first extract the PRs related to documentation changes from AIDev’s \code{pull\_request} (33,596 PRs from 2,807 repos) and \code{human\_pull\_request} (6,618 PRs from 818 repos) tables, which store PR-level metadata.
To ensure a fair comparison between Agentic-PRs and Human-PRs under the same repository conditions, we restrict Agentic-PRs to PRs from repositories with more than 500 stars (1,478 repos); the Human-PRs subset in AIDev is already filtered to the same threshold.
As a result, we obtained 1,478 documentation\hyp{}related Agentic-PRs and 519 documentation\hyp{}related Human-PRs.

Next, we retrieve the commit-level information for each collected PR.
Commit-level details for Human-PRs are not included in AIDev, so we retrieved them via the GitHub API. To ensure a consistent data collection procedure and a fair comparison, we also fetched commit-level details for Agentic-PRs via the GitHub API, even though they are available in AIDev.

For each PR, we collect commit metadata, the total number of added and deleted lines across the PR, and per-commit file-level change information, including file names, change types, and the numbers of added and deleted lines for each modified file.
From the Agentic-PRs, we retrieved 3,653 commits and 35,428 file-change records; from the Human-PRs, we retrieved 1,889 commits and 17,013 file-change records.
We additionally constructed a subset of documentation\hyp{}related PRs that actually modify documentation files.
Following the extension- and path-based heuristic used in prior work~\cite{bogomolov2024long}, we define documentation files as those with extensions such as \code{.md} and \code{.txt}, or whose paths contain tokens like \code{/docs/} or \code{README}.
From the Agentic-PRs, we retrieved 2,363 commits that include documentation-file changes and 14,410 documentation-file change records; from the Human-PRs, we retrieved 993 such commits and 4,236 documentation-file change records.

\subsection{Analysis for RQ1 (Prevalence)}

To compare documentation contributions between agentic and human developers, we first examine the number of PRs created by each group. Using the documentation\hyp{}related PRs filtered from the AIDev, we separate Agentic-PRs and Human-PRs; for Agentic-PRs, we further split them by agent type and compare PR counts across agents and against human developers.

We also analyze how agentic and human developers collaborate at the file level using all file-change records in our constructed dataset. We group commits by file path and count, for each file, whether all commits are made exclusively by one group or whether the file is co-edited by both groups through commits from agentic and human developers.

Moreover, to assess the extent to which PRs actually modify documentation, we analyze the types of files included in the commits of each PR.
In this analysis, we categorize files into ``documentation'' (``docs'') and ``non-documentation'' (``non-docs'').
Following existing research~\cite{bogomolov2024long}, we treat files with extensions such as \code{.md} and \code{.txt}, or whose paths contain tokens like \code{/docs/} or \code{README}, as ``docs'', and define all other files as ``non-docs''. For each PR, we inspect the modified files and determine whether the PR changes docs only, non-docs only, or both.

\subsection{Analysis for RQ2 (Integration)}

To examine how often AI’s documentation\hyp{}related changes are accepted, we inspect each file’s commit history and extract cases where an agent commit is followed by the file’s next human commit.
After applying this filtering, we identify 119 such human\hyp{}after\hyp{}agent cases. For each such commit, we measure the number of lines added by the agent and the number of lines deleted by the human. If the human deletes a large number of lines relative to the agent’s additions, we can infer that the agent’s changes were largely removed.

\section{Results}

\subsection{Results for RQ1 (Prevalence)}

\begin{figure}[ht]
  \centering
  \includegraphics[width=\linewidth]{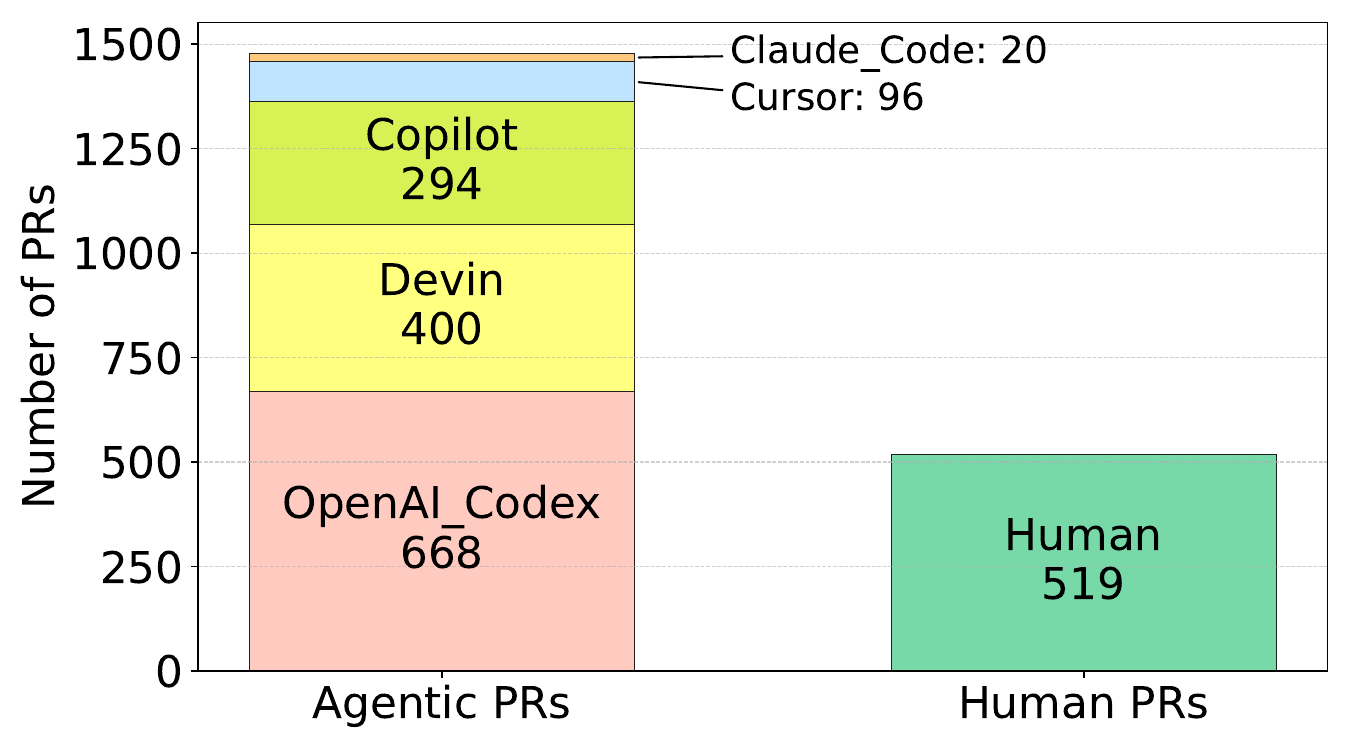}
  \caption{Number of docs-related PRs}
  \label{fig:docs_pr_counts}
\end{figure}

We show the number of documentation\hyp{}related PRs in the \figref{fig:docs_pr_counts}.
Agentic PRs (1,478) substantially outnumber Human PRs (519), indicating that AI agents are widely used even for documentation edits. Within the Agentic PRs, \code{OpenAI\_Codex} accounts for 45\%, a distribution that closely mirrors the breakdown observed across all PRs (not limited to documentation) created by AI agents~\cite{li2025aidev}.

\begin{figure}[ht]
  \centering
  \includegraphics[width=\linewidth]{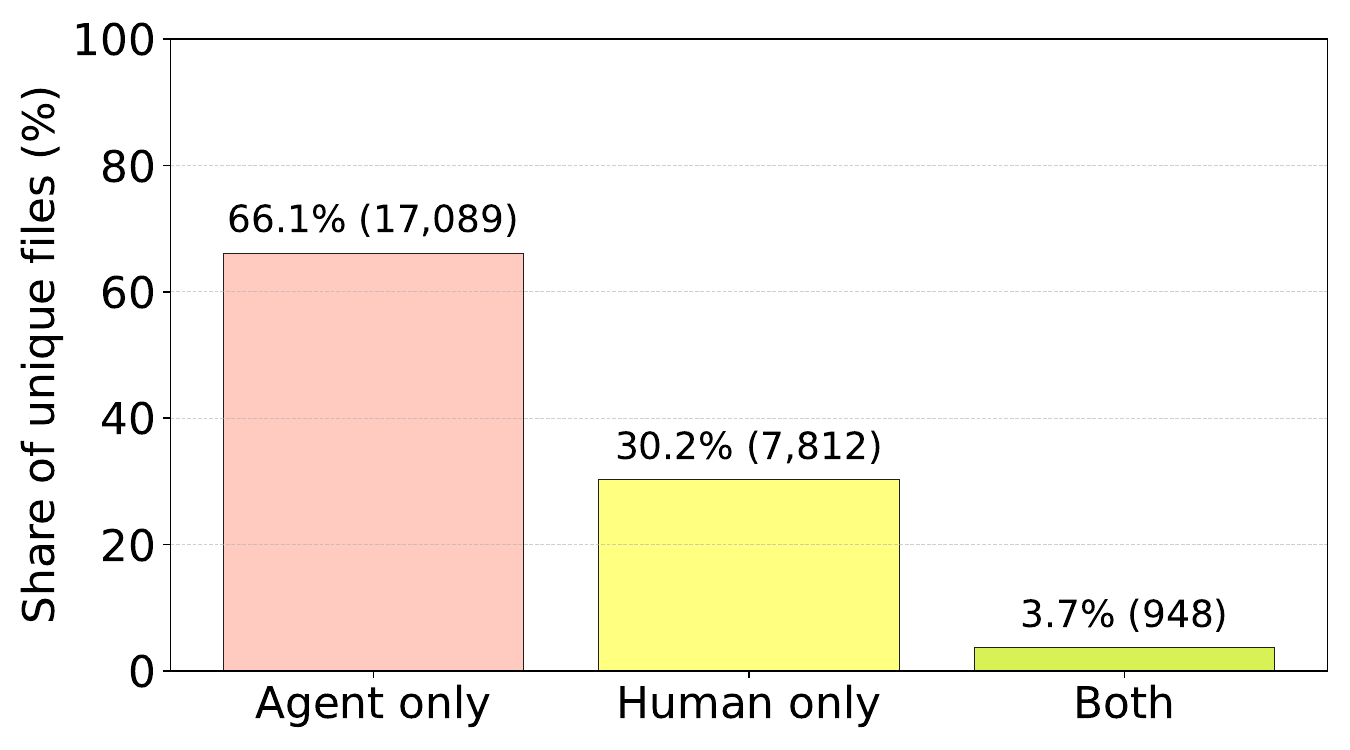}
  \caption{Breakdown of contributions by file}
  \label{fig:file_contribution}
\end{figure}

We also show the breakdown of contributions by file in \figref{fig:file_contribution}.
Only 3.7\% of files (948) are co-edited by both agents and humans, indicating that most (96.3\%) files are edited exclusively by either agents or humans rather than collaboratively.

\begin{figure}[ht]
  \centering
  \includegraphics[width=\linewidth]{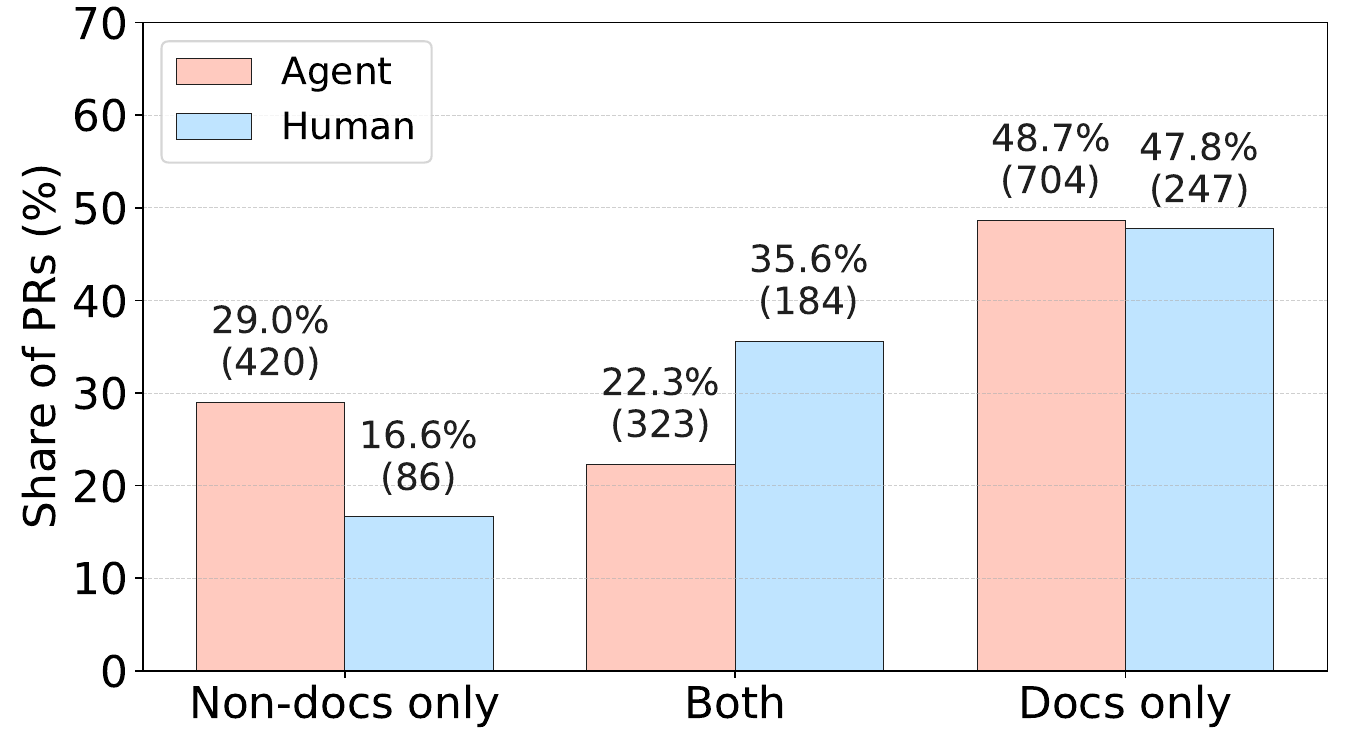}
  \caption{Distribution of PRs categorized by the types of files}
  \label{fig:pr_file_type_distribution}
\end{figure}

\figref{fig:pr_file_type_distribution} shows the distribution of PRs categorized by the types of files they include.
Among documentation\hyp{}related PRs, the share that modifies only documentation is comparable between agents and humans (48.7\% and 47.8\%, respectively).
In addition, a substantial fraction of documentation\hyp{}related PRs modifies only ``non-docs''-—i.e., does not edit documentation at all—accounting for 29.0\% and 16.6\% for agent and human PRs, respectively.
In conventional commit conventions, the \code{docs} type is intended for \emph{documentation-only} changes~\cite{oep51_conventional_commits,commitlint_conventional_types};
therefore, PRs labeled as \code{docs} that also modify non-doc files violate this expectation; such mixed-scope PRs are associated with higher reviewer effort and a greater risk that changes are overlooked, which can reduce development productivity.

\begin{tcolorbox}[colback=gray!5,colframe=awesome,title=RQ1 Summary]
Answering \textbf{RQ1}, we find that AI agents are widely used for documentation edits and that 96.3\% of
edited files are
handled exclusively by either agents or humans;
moreover,
29.0\% of agent-authored documentation-related PRs do not edit any documentation files, suggesting non-compliance with commit conventions.
\end{tcolorbox}

\subsection{Results for RQ2 (Integration)}

\begin{figure}[ht]
  \centering
  \includegraphics[width=\linewidth]{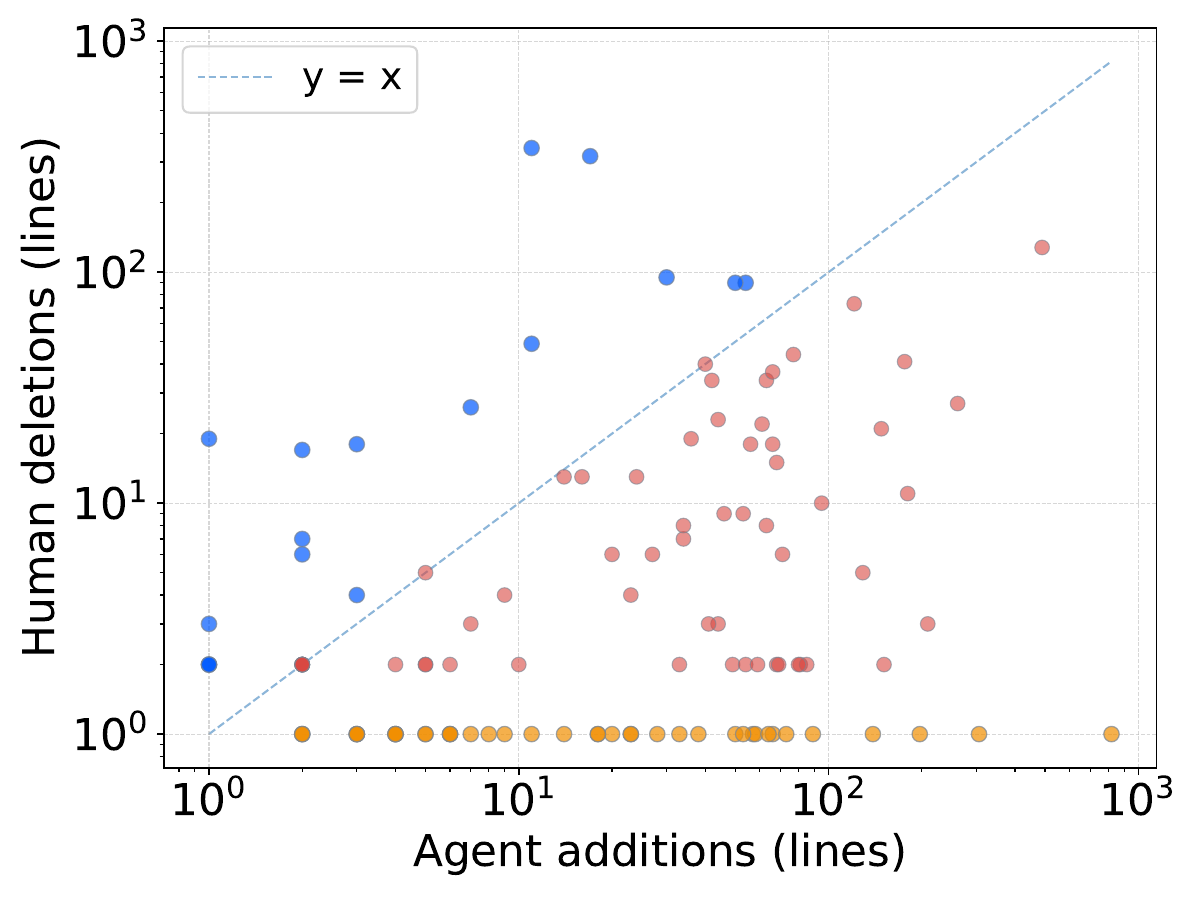}
  \caption{Human Followup Impact}
  \label{fig:human_followup_impact}
\end{figure}

\figref{fig:human_followup_impact} shows a scatter plot (log scale) of the number of lines added by agents versus the number of lines deleted by humans.
Across 119 files, we observe 17 cases (14.3\%) where human deletions exceed agent additions, and 102 cases (85.7\%) where agent additions exceed or match human deletions.
Notably, within the latter group, there are 41 files (34.5\% overall) in which humans perform no deletions at all after agent additions.
Moreover, when agent additions exceed human deletions, most added lines remain in the integrated version: on average, 86.8\% of the added lines persist (median 98.7\%).
Taken together, these results suggest that agent-authored documentation changes tend to be accepted by human developers, yet a substantial subset appears to be merged with minimal follow-up, raising questions about the level of scrutiny that agent edits receive.

\begin{tcolorbox}[colback=gray!5,colframe=awesome,title=RQ2 Summary]
Answering \textbf{RQ2}, agent-authored documentation changes are largely retained, with limited human follow-up edits.
Agent additions exceed or match human deletions in 85.7\% of files, including 34.5\% with \emph{no} human deletions.
When agent additions dominate, most added lines remain (mean 86.8\%, median 98.7\%).
\end{tcolorbox}

\section{Discussion and Future Work}

This paper makes three contributions toward understanding documentation work in SE 3.0: (1) using AIDev, we conduct a large-scale comparison showing that agents drive the volume of documentation-related PRs and that file co-editing between agents and humans is relatively rare; (2) we analyze how documentation changes are packaged in practice, showing that agent “docs”-related PRs more often include non-docs files and that some agent-authored edits are accepted with little observable human follow-up, raising review and quality-assurance concerns; and (3) we release a replication package with the analysis data and scripts to support reproducibility and follow-up studies.

We discuss three threats to validity.

For external validity, our results are limited by the set of repositories covered by the AIDev~\cite{li2025aidev}. AIDev is constructed only from repositories that contain agent-authored PRs; although we extract Agentic-PRs and Human-PRs under identical repository conditions for a fair comparison, it is important to note that repositories that do not use agents at all are not represented in the dataset.

For internal validity, our identification of docs versus non-docs files has limitations. Following existing research~\cite{bogomolov2024long}, we classify documentation files using a lightweight heuristic based on file extensions and paths; however, some PRs labeled as documentation-related in AIDev do not actually modify documentation files (e.g., 29.0\% of agent-authored documentation-related PRs touch only non-docs files). This mismatch introduces uncertainty into our estimates and interpretation of documentation work, motivating future work to better align task labels, file classification, and actual documentation edits in practice.

Moreover, our interpretation of “limited human follow-up” in RQ2 relies on quantitative signals (e.g., no human deletions and high line-retention) and does not yet consider human additions, modifications, or review discussions. Future work will move beyond aggregate metrics by examining cases with little or no follow-up (e.g., review comments, approval patterns, and subsequent edits) to assess how labels, file-level changes, and review practices align in practice.

\begin{acks}
This work has been supported by JSPS KAKENHI No. JP24K14895.
\end{acks}

\bibliographystyle{ACM-Reference-Format}
\balance
\bibliography{reference}

\end{document}